\documentclass[aps,pra,twocolumn,amsmath,amssymb,nofootinbib,superscriptaddress]{revtex4}

\newcommand{\bra}[1]{\langle#1|}
\newcommand{\ket}[1]{|#1\rangle}

\newcommand{\overlap}[2]{\langle{#1}|{#2}\rangle}

\newcommand{\BS}{{\sc BosonSampling}}

\usepackage[pdftex]{graphicx}
\usepackage{mathrsfs}
\usepackage[usenames,dvipsnames]{color}
\usepackage{hyperref}
\hypersetup{colorlinks = true, linkcolor = [rgb]{0.9059,0.4902,0.2275}, urlcolor = [rgb]{0.3882,0.3922,0.4}, citecolor = [rgb]{0.2313,0.6431,0.8824}}
\usepackage{subfig}
\usepackage{amsmath, amssymb}
\usepackage{mathtools}
\usepackage{lipsum} %for 1 column equations in 2 column manuscripts

\usepackage{ragged2e}
\DeclareCaptionJustification{justified}{\justifying}
\begin{document}

\bibliographystyle{apsrev}

\title{Implementing Scalable Boson Sampling with Time-Bin Encoding: \\ Analysis of Loss, Mode Mismatch, and Time Jitter}

\author{Keith R. Motes}
\email[]{motesk@gmail.com}
\affiliation{Centre for Engineered Quantum Systems, Department of Physics and Astronomy, Macquarie University, Sydney NSW 2113, Australia}

\author{Jonathan P. Dowling}
\affiliation{Hearne Institute for Theoretical Physics and Department of Physics \& Astronomy, Louisiana State University, Baton Rouge, LA 70803}

\author{Alexei Gilchrist}
\affiliation{Centre for Engineered Quantum Systems, Department of Physics and Astronomy, Macquarie University, Sydney NSW 2113, Australia}

\author{Peter P. Rohde}
\email[]{dr.rohde@gmail.com}
\homepage{http://www.peterrohde.org}

\affiliation{Centre for Quantum Computation and Intelligent Systems (QCIS), Faculty of Engineering \& Information Technology, University of Technology Sydney, NSW 2007, Australia}

\affiliation{Centre for Engineered Quantum Systems, Department of Physics and Astronomy, Macquarie University, Sydney NSW 2113, Australia}
\date{\today}

\frenchspacing

% ABSTRACT
\begin{abstract}
It was recently shown by Motes, Gilchrist, Dowling \& Rohde [PRL 113, 120501 (2014)] that a time-bin encoded fiber-loop architecture can implement an arbitrary passive linear optics transformation. This was shown in the case of an ideal scheme whereby the architecture has no sources of error. In any realistic implementation, however, physical errors are present, which corrupt the output of the transformation. We investigate the dominant sources of error in this architecture --- loss and mode-mismatch --- and consider how it affects the {\BS} protocol, a key application for passive linear optics. For our loss analysis we consider two major components that contribute to loss --- fiber and switches --- and calculate how this affects the success probability and fidelity of the device. Interestingly, we find that errors due to loss are not uniform (unique to time-bin encoding), which asymmetrically biases the implemented unitary. Thus, loss necessarily limits the class of unitaries that may be implemented, and therefore future implementations must prioritise minimising loss rates if arbitrary unitaries are to be implemented. Our formalism for mode-mismatch is generlized to account for various phenomenon that may cause mode-mismatch, but we focus on two --- errors in fiber-loop lengths, and time-jitter of the photon source. These results provide a guideline for how well future experimental implementations might perform in light of these error mechanisms.
\end{abstract}

\maketitle

\section{Introduction}
In 2011 a simple model of quantum simulation using linear optics -- {\BS} -- was introduced by Aaronson \& Arkhipov (AA) \cite{bib:AaronsonArkhipov10}. This protocol is not believed to be capable of universal quantum computing as was shown possible in the seminal result by Knill, Laflamme \& Milburn (KLM) \cite{bib:KLM01} in 2001. Rather, {\BS} can implement a subset of KLM linear optics quantum computing (LOQC) that does not require feedforward, quantum memory, and dynamic control, only requiring single-photon Fock state preparation, passive linear optics, and photo-detection. {\BS} has received much interest owing to its relative experimental simplicity as compared to universal quantum computing, but is nonetheless believed to implement a classically hard problem. Several elementary experimental demonstrations of {\BS} have recently been performed \cite{bib:Peruzzo17092010, bib:Broome2012, bib:Tillmann4, bib:Crespi3, bib:Spring2}. A major interest of late is to consider what other quantum states of light yield computationally interesting sampling problems that, like {\BS}, yield a classically hard problem \cite{bib:olson2014boson, bib:olson2014sampling, bib:catSampling, bib:LundScatter13}.

{\BS} is reminiscent of the Deutsch-Jozsa algorithm \cite{bib:deutsch1992rapid} of 1992. This algorithm demonstrated an exponential separation between the complexity of classical and quantum algorithms, but it solved a problem of no practical interest. Later, however, Shor described an efficient quantum factoring algorithm \cite{bib:Shor97}, a problem that is believed to be classically hard, generating much interest in the prospect of scalable quantum computing. Recently, and almost simultaneously, two separate works showed practical applications for {\BS}. It was shown by Motes, Olson, Rabeux, Dowling, Olson \& Rohde (MORDOR) that it may used for sub-shot-noise limited metrology \cite{bib:MORDOR}, and by Huh \emph{et al.} that it can be used to generate molecular vibrational spectra \cite{bib:huh2014boson}. These applications are analogous too the Shor's algorithm of the {\BS} problem --- a first glimpse into the potential of simple quantum sampling problems. For a more detailed introduction to {\BS} see Ref. \cite{bib:gard2014introduction}. Although the fiber-loop scheme was initially presented for the purposes of {\BS}, Rohde recently demonstrated that with minor modifications the scheme can be made universal for quantum computing \cite{bib:RohdeUniversal}. Here, however, we will focus on the application of this scheme to {\BS}, or purely passive linear optics applications more generally.

Although {\BS} is much easier to implement than universal LOQC, it remains experimentally challenging. The main challenges are preparing the input state and implementing the required unitary. It is possible to prepare the desired Fock state input using spontaneous parametric down-conversion (SPDC) single-photon sources \cite{bib:Motes13, bib:LundScatter13}. Loss in the unitary transformation is a problem but there is evidence that even lossy systems or systems with mode-mismatch are still likely hard to simulate given that the errors are sufficiently small \cite{bib:RohdeRalphErrBS, bib:RohdeLowFid12}. Constructing the required linear optics interferometer is challenging, as a {\BS} device might require thousands of optical elements, which must all be simultaneously aligned. Two demonstrated ways to overcome the alignment problem are to use the time-bin encoded scheme by Motes, Gilchrist, Dowling \& Rohde (MGDR) \cite{bib:PRLFiberLoop} or time-dependent dispersion techniques as presented by Pant \& Englund \cite{bib:pant2015high}.  Both methods do away with the hundreds or perhaps thousands of optical elements, requiring only a single pulsed photon-source and a single time-resolved photo-detector. An attractive feature of the former architecture is that there is only a single point of interference, and may therefore be much easier to align than conventional approaches. Additionally, the experimental complexity of these schemes are fixed, irrespective of the size of the desired interferometer. In this manuscript we focus on analysing errors in the MGDR protocol to establish how well such a protocol might behave in the presence of the two dominant sources of error --- loss and mode-mismatch.

MGDR showed that, using this architecture, arbitrary linear optics transformations can be implemented on a pulse-train of photons. However, this work assumes that the architecture has no sources of error. When errors are present the scheme no longer implements an arbitrary unitary transformation, but is constrained by the error model. In this work we analyse in detail various sources of error in the MGDR protocol. We begin by reviewing the MGDR architecture and then analyse the effects of lossy elements in the architecture followed by mode-mismatch caused by imperfect fiber-loop lengths and time-jitter in the source. These analyses accommodate the dominant challenges facing future experimental implementation. 
 
%Section
\section{Fiber-loop Architecture}

\begin{figure}[!htb]
\includegraphics[width=0.8\columnwidth]{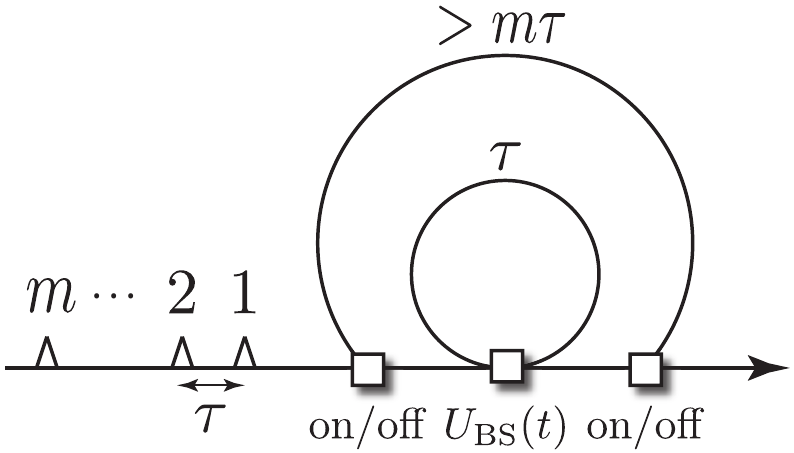}
\caption{The complete fiber-loop architecture fed by a pulse-train of photonic modes, each separated in time by $\tau$. The squares represent optical switches. A single length $\tau$ inner fiber-loop is embedded inside a length \mbox{$>m\tau$} outer fiber-loop. The outer loop allows an arbitrary number of inner loops to be applied consecutively. When \mbox{$m-1$} inner loops are implemented this architecture realises an arbitrary unitary transformation on $m$ modes given that no loss is present.} \label{fig:full_architecture}
\end{figure}

It was shown by MGDR that an arbitrary unitary can be implemented using time-bin encoding in a loop-based architecture \cite{bib:PRLFiberLoop}. This is useful for the {\BS} model because it significantly reduces the number of required optical elements. In fact, the experimental requirements to implement the architecture are fixed, irrespective of the dimension of the desired unitary. Thus, the scheme is highly scalable, and uses far fewer physical resources than schemes based on bulk-optics or integrated waveguides. In this architecture, shown in Fig. \ref{fig:full_architecture}, a pulse-train of photonic modes consisting of, in general, Fock states and vacuum, are each separated by time $\tau$ and sent into an embedded fiber-loop. The $i$th time-bin corresponds to the $i$th mode in a conventional spatially-encoded scheme. 

The boundary conditions of the protocol are that the first time-bin is coupled completely into the inner loop and the last time-bin coupled completely out of the inner loop (after it traverses the inner loop once), such that the implemented unitary is bounded as an \mbox{$m\times m$} matrix, where $m$ is the length of the pulse-train. This can be obtained with, 
\begin{equation} \label{eq:VarBSboundary}
\hat{U}_{\mathrm{BS}}\left(1\right) = \hat{U}_{\mathrm{BS}}\left(m+1\right)=\left( \begin{array}{cc}
0 & 1  \\
1 &0 
\end{array} \right),
\end{equation}
where $U_\mathrm{BS}(i)$ is the unitary associated with the central beamsplitter at time $i$. The pulse-train then evolves in the inner loop. Each mode takes time $\tau$ to traverse the inner loop so that it will interfere with the next time-bin at the central beamsplitter. Between each pulse a dynamically controlled beamsplitter, $\hat{U}_{\mathrm{BS}}(t)$ of the form, 
\begin{eqnarray} \label{eq:VarBS}
\hat{U}_{\mathrm{BS}}(t) &=& \left( \begin{array}{cc}
u_{1,1}(t) & u_{1,2}(t)  \\
u_{2,1}(t) & u_{2,2}(t) 
\end{array} \right),
\end{eqnarray}
where $\hat{U}_\mathrm{BS}$ is an arbitrary, time-dependent SU(2) operation, is applied. Here $u_{i,j}$ is the amplitude of input mode $i$ reaching output mode $j$. \mbox{$i=1$} (\mbox{$i=2$}) represents the mode entering from the source (inner loop), and \mbox{$j=1$} (\mbox{$j=2$}) represents the mode exiting the loop (entering the loop). When a mode enters the loop it progresses to the next time-bin.

After the entire pulse-train exits the inner loop the unitary map $\hat{V}$ is implemented, 
\begin{equation} \label{eq:UnitaryMap}
V_{i,j} = \left\{ \begin{array}{ll}
 0 & i>j+1 \\
 u_{1,1}(i) & i=j+1 \\
 u_{1,2}(i)u_{2,1}(j+1) \prod_{k=i+1}^{j}{u_{2,2}}(k) & i<j+1
 \end{array} \right. ,
\end{equation}
where \mbox{$i\in\{1,m\}$} and \mbox{$j\in\{1,m\}$} represent input and output modes respectively. This map may easily be seen by writing out the spatial representation of the implemented unitary map and carefully following how each input mode traverses through to each output mode as done in the original MGDR work. Note that we have employed a slightly different, but equivalent, indexing convention to the original MGDR proposal. %This representation is desirable because the matrix elements of $U_{\mathrm{BS}}$ correspond to entering the next mode if it goes into the loop and staying in the same mode if it exits the loop. Also, this representation suits the mode-mismatch calculations of Sec. \ref{sec:ModeMismatch} more appropriately.

%When $i>j+1$ the $i$th input mode does not have access to the $j$th output mode so this matrix element is zero. When $i=j+1$ the modes do not enter the inner loop and travel strait through to the detector picking up a factor of $\gamma_{2,1}(i)$. When $i<j+1$ the modes traverse the loop $j-i+1$ times.

The inner loop alone cannot implement an arbitrary unitary transformation so additional loops are required. The outer loop allows for an arbitrary number of applications of the  inner loop to be implemented. The net unitary $\hat{U}$ after $L$ consecutive inner loops becomes, 
\begin{equation}
\hat{U}=\prod_{l=1}^{L}\hat{V}(l),
\end{equation}
where $l$ denotes the $l$th iteration of the inner loop. The pulse-train will traverse the inner loop \mbox{$L=m-1$} times and the outer loop \mbox{$m-2$} times for an arbitrary unitary transformation to be implemented. The outer loop must have round trip time \mbox{$>m\tau$} so that the pulse-train does not interfere with itself for a particular instance of the inner loop $\hat{V}(l)$. The pulse-train is coupled in and out of the outer loop via on/off switches. Once the desired transformation is performed, the pulse-train exits both loops and is measured via time-resolved photo-detection, where the time-resolution of the detector must be greater than $\tau$. The $j$th time-bin at the output corresponds to the $j$th spatial mode in the standard {\BS} model. 

The unitary derived above assumes lossless components and perfect mode-matching at the central beamsplitter. In any realistic implementation this will not be the case, which we consider next.

%Section
\section{Loss Errors} \label{sec:LossErrors}

In an implementation of a passive linear optics network, whereby the loss between each input/output pair of modes is uniform, loss simply amounts to a reduced success probability upon post-selecting on detecting all photons. In the fiber-loop architecture, the different paths traverse the inner loop a different number of times leading to non-uniform loss. This biases the unitary transformation resulting in a unitary that is not the desired one, even after post-selecting upon measuring all photons. That is, the effects of loss cannot be simply factored out of the unitary. In some architectures, asymmetric losses may be compensated for by artificially adding losses that rebalance the circuit, at the expense of overall success probability. In the fiber-loop architecture this turns out not to be the case. 

In Sec. \ref{sec:LossMetrics} we introduce the metrics that we will use to analyse loss. In Sec. \ref{sec:innerLoopLoss} we  determine the effect of loss due to the lossy switch and lossy fiber in the inner loop. Then in Sec. \ref{sec:OuterLoopLoss} we analyse the net loss combining the inner loop losses with the outer loop losses. We denote quantities here that have loss with a prime.

\subsection{Loss Metrics} \label{sec:LossMetrics}

\subsubsection{Similarity}

An interesting question is how small does loss need to be such that a particular unitary transformation is implemented with a particular error bar. The answer to this question is highly dependent on which unitary we wish to implement --- some unitaries will suffer more asymmetric bias than others, depending on the switching sequence that is required to implement them. Thus, the first question to ask is which unitary to consider. In the work of MGDR, a so-called `uniform' unitary was considered. This is a unitary where the amplitude (but not necessarily phases) of each element of the unitary are equal. That is, the magnitude of the amplitude between each input/output pair of modes is the same. This class of unitaries was considered as an example of `non-trivial' matrices, which uniformly mix every input mode with every output mode. However, it is still an open question as to exactly what classes of unitaries yield hard sampling problems in the context of {\BS}. We will here consider the same setting. We will explore this by using the similarity metric, $\mathcal{S}$, which compares the implemented map with the uniform map,
\begin{eqnarray}
\mathcal{S} &=& \max_{\hat{U}_\mathrm{BS}(t) \,\forall\, t}\left[\frac{\left(\sum_{i,j=1}^{m}\sqrt{|U_{i,j}|^2 \cdot |\mathcal{W}_{i,j}|^2}\right)^2}{\left(\sum_{i,j=1}^{m}|U_{i,j}|^2\right) \cdot \left(\sum_{i,j=1}^{m}|\mathcal{W}_{i,j}|^2\right)}\right] \nonumber \\
&=&\max_{\hat{U}_\mathrm{BS}(t)\,\forall\, t}\left[\frac{1}{m^2}\frac{\left(\sum_{i,j=1}^{m}|U_{i,j}|\right)^2}{\sum_{i,j=1}^{m}|U_{i,j}|^2}\right],
\end{eqnarray}
where $\mathcal{W}_{i,j}$ is an \mbox{$m\times m$} uniform unitary given by \mbox{$|\mathcal{W}_{i,j}|^2=1/m$}. $\mathcal{S}$ is maximised by performing a Monte-Carlo simulation over different beamsplitter ratios so as to find the optimal switching sequence to make the map as uniform as possible. 

%The best possible similarity is when all of the photons exit in the same mode that they entered implementing the identity transformation. The total loss is uniform across all $U_{i,i}$ and thus factors out when the identity is implemented. Since the similarity metric renormalises $U$ then $U$
\subsubsection{Post-selection Probability}
Another interesting question is how the probability of post-selecting upon all $n$ photons is affected by loss, i.e the total success probability of the device. This is of especial importance experimentally, as it directly translates to count rates. The post-selection probability of detecting all $n$ photons at the output is,
\begin{equation} 
\mathcal{P}_{\mathrm{S}} = \prod_{i=1}^{m}\left(\sum_{j=1}^{m}|U_{i, j}|^{2}\right)^{k_i},
\label{eq:Psel}
\end{equation}
where $\{k\}$ is an integer string of length $m$ that represents a known input configuration of photons and $k_i$ is the number of photons in mode $i$. This equation is intuitively derived as follows. For a single photon the probability of entering mode $i$ and exiting mode $j$ is $|U_{i,j}|^2$. Then the total probability that the $i$th photon exits the architecture is the sum of this over all $j$ possible output ports, i.e. \mbox{$\sum_{j=1}^{m}|U_{i,j}|^2$}. Thus the probability of detecting all $n$ photons at the output beginning in a particular configuration $\{k\}$ is the product of this probability over all modes $i$ where \mbox{$k_i\neq0$}, as per Eq. \ref{eq:Psel}. This generalisation, by allowing arbitrary strings $\{k\}$, allows for implementations such as randomised {\BS} as described by Lund \emph{et al.} \cite{bib:LundScatter13}. 

With losses present, $\hat{U}$ is in general no longer unitary. Rather, it is a mapping of input-to-output amplitudes, and will not be normalised. When there is no loss in the architecture \mbox{$\mathcal{P}_\mathrm{S}=1$}, and with loss strictly \mbox{$\mathcal{P}_\mathrm{S}<1$}, dropping exponentially with the number of photons. Implementing the required \mbox{$m-1$} loops will have exponentially worse loss than a single loop. 

%\textbf{The worst case $P_S$ is when all $m-1$ loops are required and all of the photons traverse the inner loop a maximal number of times. exit the loop in mode $m$, since in this instance all photons traverse the inner loop the maximum possible number of times. Then the post-selection probability is given by $P_{\mathrm{S}} = \prod_{i=1}^{n}|U_{i, m}|^{2}$. }

%\textbf{The best case $P_S$ is when the identity is implemented. To implement the identity only one inner loop is required and each photon only traverses the inner loop once. The post-selection probability in this case is given by $P_{\mathrm{S}} = \prod_{i=1}^{n}|U_{i, i}|^{2}$.}

\subsection{Inner Loop Loss} \label{sec:innerLoopLoss}

\begin{figure}[!htb]
\includegraphics[width=.8\columnwidth]{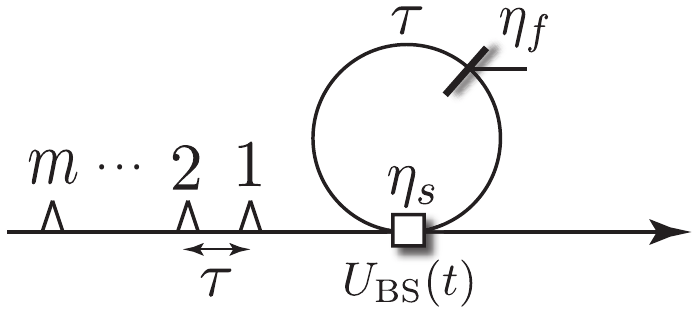}
\caption{A lossy inner fiber-loop fed by a pulse-train of photonic modes, each separated in time by $\tau$. We model the loss of the loop with a beamsplitter of reflectivity $\eta_f$ and the loss of the switch with an efficiency $\eta_s$. Each mode experiences different amounts of loss, i.e. the first mode traverses the loop up to $m$ times, the second up to \mbox{$m-1$} times, \dots, and the $m$th mode at most once.} \label{fig:singlelooperror}
\end{figure}

We will model loss inside of the inner fiber-loop with a beamsplitter of reflectivity $\eta_f$ and loss in the switch as $\eta_s$ as shown in Fig. \ref{fig:singlelooperror}. When \mbox{$\eta_f=\eta_s=1$} the device has perfect efficiency. Before and after the inner loop the loss experienced by each mode in the fiber is negligible, since it may be arbitrarily short. Taking these losses into account, the implemented map of Eq. \ref{eq:UnitaryMap} becomes,
\begin{equation} \label{eq:lossyUnitaryMap}
V'_{i,j} = \eta_s\left\{ \begin{array}{ll}
 0 & i>j+1 \\
 u_{1,1}(i) & i=j+1 \\
 \eta^{j-i+1}u_{1,2}(i)u_{2,1}(j+1)\cdot \\ \prod_{k=i+1}^{j}{u_{2,2}}(k) & i<j+1
 \end{array} \right. ,
\end{equation}
for a given loop, where \mbox{$\eta=\eta_f\eta_s$}. Note that this mapping is no longer a unitary matrix when \mbox{$\eta_f<1$} or \mbox{$\eta_s<1$}. This uneven distribution of losses in the input-to-output mapping causes a skew in the matrix which prevents it from implementing the desired unitary transformation, even after post-selection. 

\subsection{Outer Loop Loss} \label{sec:OuterLoopLoss}

\begin{figure}[!htb]
\includegraphics[width=.8\columnwidth]{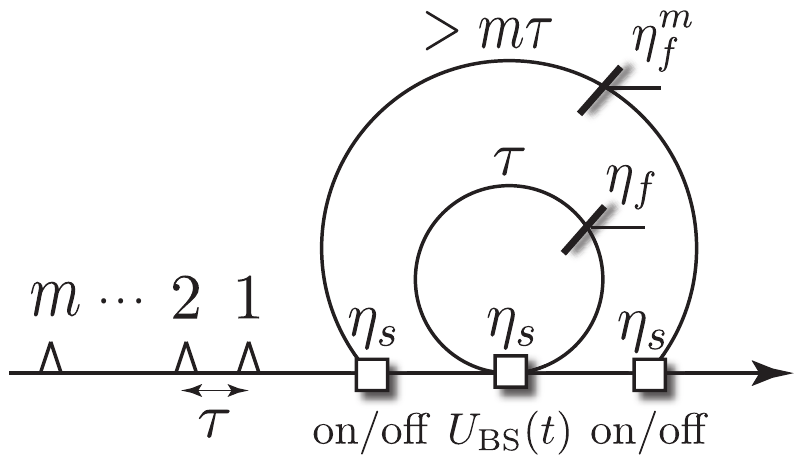}
\caption{The full architecture which implements the lossy transformation $\hat{U}'$. Each mode experiences ${\eta_f}^m$ loss per outer loop since they each take time $m\tau$ to traverse the outer loop. For an arbitrary unitary to be implemented in the ideal case the photons will traverse the outer loop $L-1$ times. This yields a net fiber loss from the outer loop of ${\eta_f}^{m(L-1)}$ that can be factored out of $\hat{U}'$, since it affects all paths equally. The net switch loss from the outer two switches is ${\eta_s}^{2(L-1)}$ and can also be factored out of $\hat{U}'$. The losses within the inner loop, on the other hand, affect different paths differently, and in general cannot be factored out.} \label{fig:FullLossError}
\end{figure}

In the full fiber-loop architecture $L$ inner loops are implemented via \mbox{$L-1$} round-trips of the outer loop, before being coupled out to the detector. This architecture can implement an arbitrary unitary transformation when \mbox{$L=m-1$} if there are no errors present. The outer loop and outer switches cause a uniform loss on the entire pulse-train, since every path through the interferometer passes through these elements the same number of times. Hence, these factor out of $\hat{U}$. The full lossy transformation that occurs is then, 
\begin{equation}
\hat{U}'={\eta_{f}}^{m(L-1)}{\eta_{s}}^{2(L-1)}\prod_{l=1}^{L}\hat{V}'(l),
\end{equation}
where $L=m-1$ if an arbitrary transformation is desired, and $\hat{V}'$ is given by Eq. \ref{eq:lossyUnitaryMap}. 
The ${\eta_{f}}^{m(L-1)}$ occurs because the pulse-train traverses an $m\tau$ length of fiber in the outer loop $L-1$ times (i.e $\eta_f$ can be regarded as the efficiency per unit of fiber of length $\tau$), and the ${\eta_{s}}^{2(L-1)}$ occurs because the pulse-train passes through the two outer switches \mbox{$L-1$} times. Fig. \ref{fig:FullLossError} shows the entire architecture with these loss errors. For an example of loop bias due to loss see App. \ref{app:MultipleLoopBiasExample}. Extending from this loop bias example we generalize the loss matrix denoted as $\hat{\mathcal{L}}$, which represents the accumulation of losses in the fiber-loop architecture, as a function of the number of loops $L$ for an arbitrarily sized \mbox{$m\times m$} transformation,
\begin{equation} \label{eq:lossyMatrix}
\mathcal{L}_{i,j}(L) = \eta_s^L \eta^{L+j-i},
\end{equation}
again where $\eta=\eta_f\eta_s$. Now the lossy map $\hat{U}'$ may be written as an element wise product of the ideal unitary $\hat{U}$ and the loss matrix $\hat{\mathcal{L}}$,
\begin{eqnarray}
\hat{U}' &=& \hat{U}\circ\hat{\mathcal{L}}.
\end{eqnarray}
Elements of $\hat{\mathcal{L}}$ that have no losses in them due to input modes not reaching output modes when $L<m-1$ will be accounted for appropriately when $\hat{\mathcal{L}}$ is multiplied by $\hat{U}$ by making the cooresponding matrix element in $\hat{U}'$ go to zero.

\begin{figure}[!htb]
\centering
\subfloat[Part 1][]
{\includegraphics[width=.8\columnwidth]{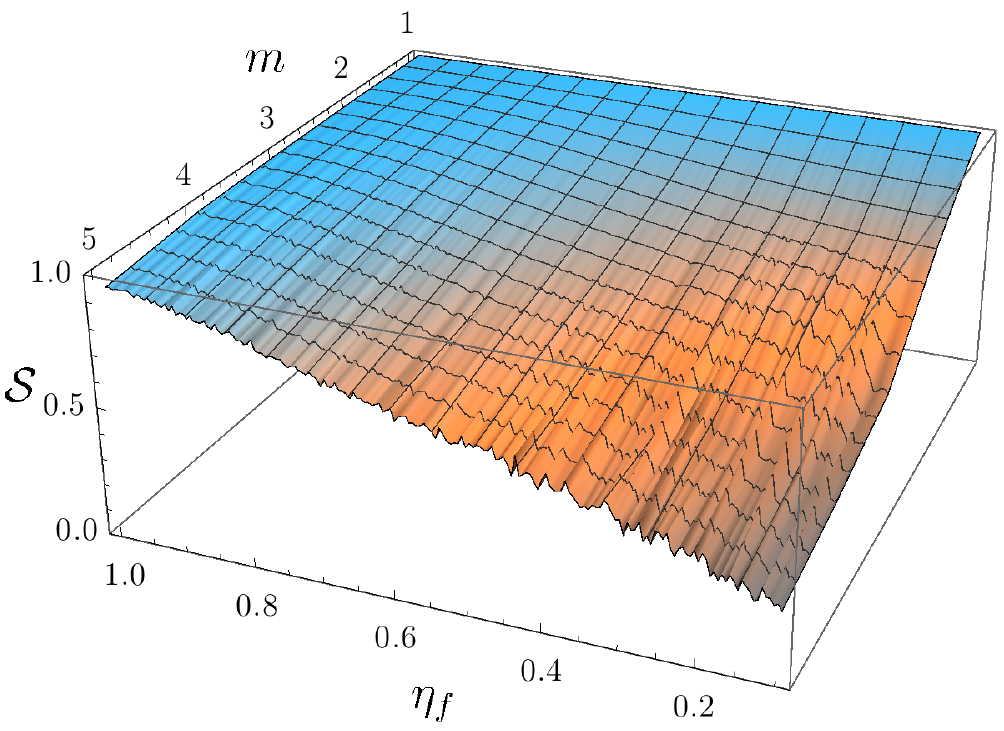} \label{fig:SL}}\\
\subfloat[Part 2][]
{\includegraphics[width=.8\columnwidth]{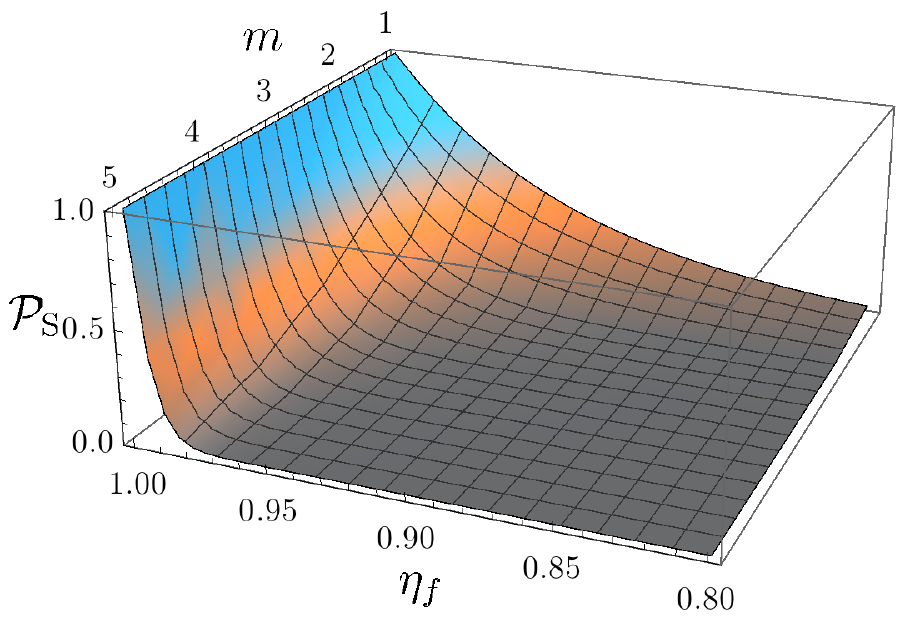} \label{fig:PSL}}
\caption{(a) Similarity $\mathcal{S}$ versus mode/photon number \mbox{$m=n$} and loop efficiency $\eta_f$ for \mbox{$m-1$} loops. The map remains similar to the uniform unitary for low loss rates, implying that non-trivial unitary transformations may be implemented. \mbox{$m-1$} loops are considered because this is the number of loops required to implement an arbitrary unitary transformation in the lossless case. (b) Post-selection probability $\mathcal{P}_\mathrm{S}$ versus mode/photon number \mbox{$m=n$} and loop efficiency $\eta_f$ for \mbox{$m-1$} loops. These two plots are related in that each point in $\mathcal{P}_\mathrm{S}$ was calculated from the switching sequence $\hat{U}'$ corresponding to that which maximises $\mathcal{S}$. In both (a) and (b) the data was averaged over 1750 Monte-Carlo iterations and we let \mbox{$\eta_s=1$}, i.e the switches are ideal but the fibers are not.}
\label{fig:SPSL}
\end{figure}

In Fig. \ref{fig:SL} we show how the optimised similarity with the uniform distribution varies with $\eta_f$ and $m$ for \mbox{$L=m-1$} inner loops, one photon in all $m$ modes, and \mbox{$\eta_s=1$}. With low loss rates (\mbox{$\eta_f \approx 1$}) the implemented unitary remains highly uniform. However, with several loops the success probability of detecting all $n$ photons at the output decays exponentially as shown is Fig. \ref{fig:PSL}. For these plots the randomly generated $\hat{U}'$ that maximises $\mathcal{S}$ for each $\eta_f$ and $m$ is used to calculate the corresponding $\mathcal{P}_\mathrm{S}$. 

\begin{figure}[!htb]
\centering
\subfloat[Part 1][]
{\includegraphics[width=.8\columnwidth]{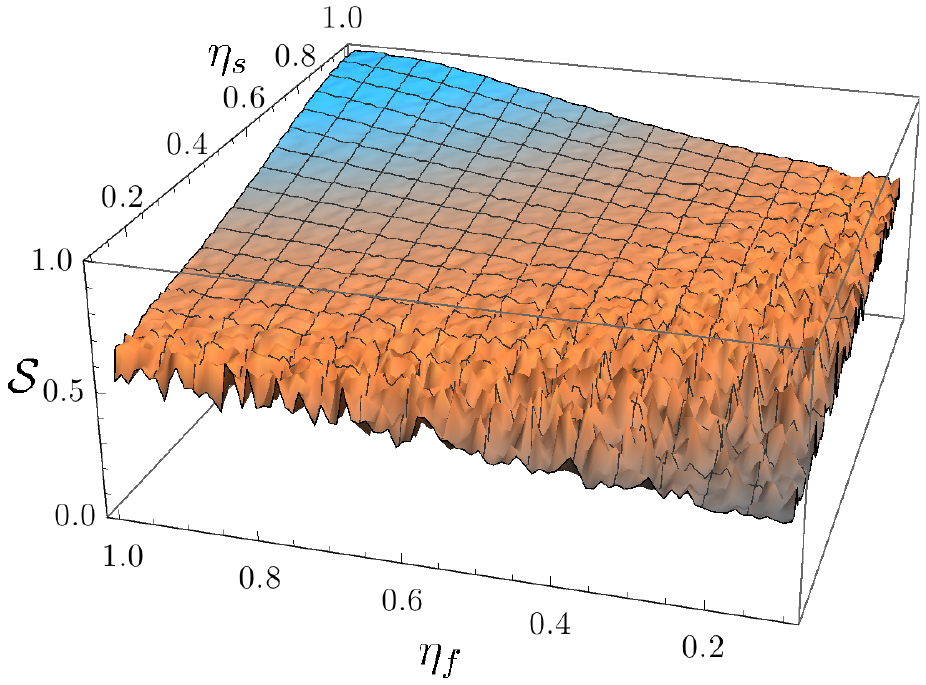} \label{fig:SLL}}\\
\subfloat[Part 2][]
{\includegraphics[width=.8\columnwidth]{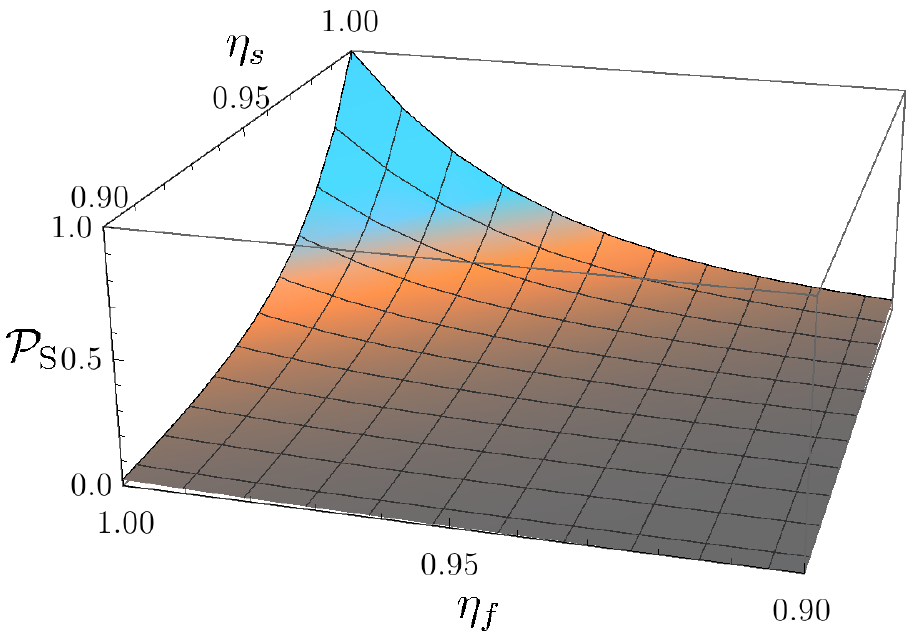} \label{fig:PSLL}}
\caption{(a) Similarity $\mathcal{S}$, and (b) post-selection probability $\mathcal{P}_\mathrm{S}$ versus loop efficiency $\eta_f$ and switch efficiency $\eta_s$ with \mbox{$m=3$} modes, one photon per input mode, and \mbox{$m-1$} loops. These two plots are again related in that $\mathcal{P}_\mathrm{S}$ is calculated from the switching sequence that maximises $\mathcal{S}$. This data was averaged over 1750 iterations.}
\label{fig:SPSLL}
\end{figure}

Now we consider how $\mathcal{S}$ and $\mathcal{P}_\mathrm{S}$ are affected in Fig. \ref{fig:SPSLL} with both the fiber loss and switch loss. We show this for the case of \mbox{$m=3$} and one photon per input mode, which is in the regime of present-day demonstrations. 
%Section

\section{Mode-matching Errors} \label{sec:ModeMismatch}

In any interferometric experiment it is inevitable that mode-mismatch will occur and is thus an essential source of error that we will consider in this section. There are many factors that may contribute to mode-mismatch in this architecture, such as incorrect fiber lengths, time-jitter in the sources, beamsplitter misalignment, and dispersion of the wave-packets. In this section we will focus on two major sources of mode-mismatch: incorrect fiber lengths and source time-jitter. The former results in reduced Hong-Ou-Mandel visibility at the central beamsplitter, owing to mismatched arrival times of photons. The latter effectively results in randomisation of the preparation times of the photons.

We consider how mode-mismatch affects our protocol by calculating the fidelity, $\mathcal{F}$, between the ideal output state $\ket{\psi_i}$ that one expects theoretically with no errors present, and the actual experimentally obtained output state $\ket{\psi_a}$. Imperfect fiber lengths and time-jitter both cause temporal shifts in the centre of the wave-packet, which will affect the output by both introducing uncertainty into the timing of the bins reaching the detector, and undermining the Hong-Ou-Mandel visibility at the central beamsplitter. To calculate $\mathcal{F}$ then we need to calculate the temporal overlap between $\ket{\psi_i}$ and $\ket{\psi_a}$. Therefore, we need to consider the temporal structure of the photons.

We will model the temporal structure of photons using the formalism of Rohde \emph{et al.} \cite{bib:spectralStructure}. We only consider the inner loop in this analysis because there is no interference at any point in the outer loop. We obtain lower and upper bounds on $\mathcal{F}$ by performing a Monte-Carlo search over different randomly generated unitaries $V$. We could also instead consider $V'$ in this formalism to also jointly include losses. But we will treat losses separately from mode-mismatch for simplicity.

\subsection{Temporal Structure of Photons}

The temporal structure of a photon can be represented using a mode operator,
\begin{equation} \label{eq:TPCO}
\hat{\mathcal{A}}^{\dag}(t,\Delta)=\int_{-\infty}^{\infty}\psi(x-t-\Delta)\hat{a}^{\dag}(x)dx,
\end{equation}
where \mbox{$\psi(t-\Delta)$} is the temporal density function centered at time $t$, $\Delta$ is a shift of the temporal centre of the photon, and $\hat{a}^{\dag}(t)$ is the time-dependent photon creation operator. This operator \mbox{$\hat{\mathcal{A}}^{\dag}(t,\Delta)$} acts on the vacuum $\ket{0}$ to create a photon with normalised Gaussian spectral density function,
\begin{equation} \label{eq:psi}
\psi(x)= \frac{1}{\sqrt{c\sqrt{\pi}}} e^{-\frac{x^2}{2c^2}},
\end{equation}
where $c/\sqrt{2}$ is the standard deviation. We assume that \mbox{$\tau\gg\Delta$}, in which case $t$ denotes a time-bin, and $\Delta$ denotes a small mismatch within the respective time-bin, not large enough to cause a photon to `jump' from one time-bin to the next. Thus, both $t$ and $\Delta$ represent shifts in the centre of the photon's wavepacket, but the former is of the order of the time-bin separation, while the latter is of much smaller order than the time-bin separation.

\vspace{2mm}
\subsection{Our Formalism}
\begin{figure}[!htb]
\includegraphics[width=.8\columnwidth]{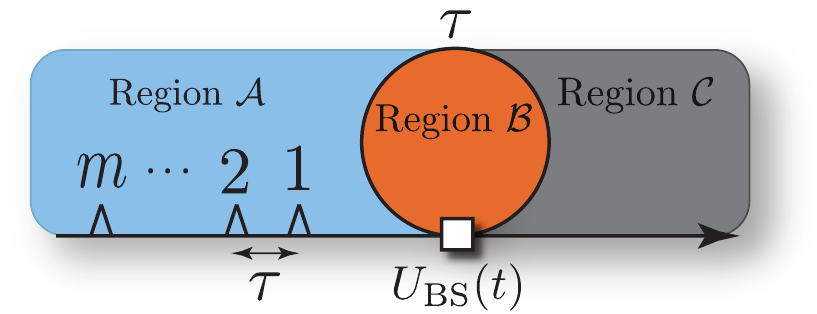}
\caption{The three regions we consider in the mode-mismatch formalism. Region $\mathcal{A}$ corresponds to the modes coming from the source, region $\mathcal{B}$ to the modes inside the inner loop, and region $\mathcal{C}$ to the modes exiting the loop.} \label{fig:Region}
\end{figure}

To analyse mode-mismatch we will consider three regions of the architecture we label as $\mathcal{A}$, $\mathcal{B}$, and $\mathcal{C}$ as shown in Fig. \ref{fig:Region}. Region $\mathcal{A}$ corresponds to the modes that are input into the architecture from the source, region $\mathcal{B}$ corresponds to pulses inside the inner loop, and region $\mathcal{C}$ corresponds to pulses that exit the dynamic beamsplitter towards the detector. We introduce mode operators associated with each of these distinct regions --- \mbox{$\hat{\mathcal{A}}^{\dag}(t,\Delta)$}, \mbox{$\hat{\mathcal{B}}^{\dag}(t,\Delta)$}, and \mbox{$\hat{\mathcal{C}}^{\dag}(t,\Delta)$} --- each of the form of Eq. \ref{eq:TPCO}.  

Since every pulse begins in region $\mathcal{A}$ the input state is a tensor product of pure states of the form,
\begin{equation} \label{eq:inputState}
\ket{\Psi_\mathrm{in}}= \bigotimes_{i=1}^{m} \frac{1}{\sqrt{k_i!}}\hat{\mathcal{A}}^{\dag}(t_i,\Delta_i)^{k_i}\ket{0}_i,
\end{equation} 
where the tensor product is taken over all $m$ modes, $\{k\}$ is a known string representing the input photon-number configuration, and $k_i$ is number of photons in the $i$th input mode.

Next, the input state is transformed by the dynamic beamsplitter, which takes the mode-operators from region $\mathcal{A}$ into superpositions of regions $\mathcal{B}$ and $\mathcal{C}$, 
\begin{eqnarray} \label{eq:BSA}
\hat{U}_{\mathrm{BS}}(t)\hat{\mathcal{A}}^{\dag}(t,\Delta)\hat{U}_{\mathrm{BS}}^{\dag}(t)&\to& u_{1,2}(t)\hat{\mathcal{B}}^{\dag}(t+1,\Delta) \nonumber \\
&+& u_{1,1}(t)\hat{\mathcal{C}}^{\dag}(t,\Delta),
\end{eqnarray}
and pulses from region $\mathcal{B}$ to superpositions of regions $\mathcal{B}$ and $\mathcal{C}$, 
\begin{eqnarray} \label{eq:BSB}
\hat{U}_{\mathrm{BS}}(t)\hat{\mathcal{B}}^{\dag}(t,\Delta)\hat{U}_{\mathrm{BS}}^{\dag}(t)&\to& u_{2,2}(t)\hat{\mathcal{B}}^{\dag}(t+1,\Delta) \nonumber \\
&+& u_{2,1}(t)\hat{\mathcal{C}}^{\dag}(t,\Delta),
\end{eqnarray}
where we have used Eq. \ref{eq:VarBS} for the elements of the dynamic beamsplitter at time $t$. $\hat{U}_{\mathrm{BS}}(t)$ only acts on photons arriving at the beamsplitter at time \mbox{$t\pm\Delta$} since \mbox{$\tau \gg \Delta$}. When a photon enters the loop \mbox{$t\to t+1$} as it advances to the next time-bin and will interfere with the next temporal mode. After this evolution, the entire pulse-train is coupled out of the loop such that the entire output state is a superposition of all possible output configurations.

Now we model the state of the pulse train after $t$ beam-splitters have been implemented,
\begin{widetext}
\begin{eqnarray} \label{eq:outevo}
\ket{\Psi(t)} &=& \left[\prod_{i'=1}^{t}\hat{U}_{\mathrm{BS}}(i')\right]\cdot \ket{\Psi_\mathrm{in}}  \nonumber \\
&=& \left[\prod_{i'=1}^{t}\hat{U}_{\mathrm{BS}}(i')\right] \cdot \left[\prod_{i=1}^{m} \frac{1}{\sqrt{k_i!}}\hat{\mathcal{A}}^{\dag}(t_{i},\Delta_i)^{k_i}\right]\cdot \left[\prod_{i'=1}^{t}\hat{U}_{\mathrm{BS}}(i')\right]^\dag \ket{0}^{\otimes m}
\end{eqnarray}
\end{widetext}
where the integer values of $t$ denote the distinct time-bins. We note that there are $m+1$ total beam-splitters in a single implementation of the inner loop since there are \mbox{$m-1$} beamsplitters to interfere the modes and another two beamsplitters to account for the initial and final boundary conditions of the MGDR protocol. Given how we modelled how the mode operators are transformed by $\hat{U}_{\mathrm{BS}}(t)$ in Eqs. \ref{eq:BSA} and \ref{eq:BSB} the $t$th beam-splitter acts on the mode operators only in modal position $t$. Since a pulse coming out of the inner loop exits at beam-splitter $t$ its modal position is $m=t-1$ which accounts for there being $m+1$ beam-splitters and $m$ modes.

In general the final evaluated form of $\ket{\Psi_\mathrm{out}}$ may be expressed as a superposition of all possible output photon-number configurations $S$, and their associated temporal configurations $T(S)$,
\begin{equation}
\ket{\Psi_\mathrm{out}}= \sum_{S}\sum_{T(S)}\bigg[\gamma_{S,T}\prod_{i=1}^{n} \hat{\mathcal{C}}^{\dag}\left(t_{S_i},\Delta_{T(S_i)}\right)\bigg] \ket{0}^{\otimes m},
\end{equation}
where $\gamma_{S,T}$ is the probability amplitude associated with photon time-bin configuration $S$ and temporal shift configuration $T(S)$, $t_{S_i}$ denotes the time-bin of the $i$th photon, $T(S)$ denotes a configuration of temporal shifts associated with the configuration $S$, and $\Delta_{T(S_i)}$ is the temporal shift of the $i$th photon associated with configurations $S$ and $T$. This is the most general representation of a configuration of photons across time-bins with associated shifts. The probability of measuring a particular configuration is $|\gamma_{S,T}|^2$, and to evaluate these probabilities we must fully characterise spectrum of time-bin and temporal shift configurations, $S$ and $T$. Finding analytic forms for these expressions is largely prohibitive, and we calculate the $\gamma_{S,T}$ via brute-force simulation of the evolution of the mode-operators through the network as described earlier.

%\begin{comment}%%%%
% If every input mode had access to every output mode, then the total number of possible configurations $|S|$ would be $m^n$. This is not the case since we are only accounting for one instance of the inner loop. We find $|S|$ by realising that the $i$th input mode only has access to the output modes \mbox{$j\geq i-1\ \forall\ j>2$},
%\begin{equation}
%|S|=m^{k_1+k_2}\prod_{i=1}^{m-2}(m-i)^{k_{i+2}}.
%\end{equation}
%\end{comment}%%%%

%%%
%\begin{comment}
%\begin{equation}
%\gamma_S = \frac{\mathrm{perm}(V_{T,S})}{\prod_{i=1}^{m}\sqrt{T_i!S_i!}},
%\end{equation}
%where $V_{T,S}'$ is the sub-matrix of the lossy input to output map associated with input configuration $T$ and output configuration $S$. 
%\end{comment}
%%%

\subsection{Fidelity Metric}
We analyse the results of this section by calculating the fidelity $\mathcal{F}$ between the ideal output state and the actual output state, given by,
\begin{equation} 
\mathcal{F}= |\overlap{\Psi_\mathrm{i}}{\Psi_\mathrm{a}}|^2,
\end{equation}
where $\ket{\Psi_\mathrm{i}}$ is the ideal output state with no mode-mismatch (\mbox{$\Delta \to 0$}) and $\ket{\Psi_\mathrm{a}}$ is the actual output state obtained with mode-mismatch. $\ket{\Psi_\mathrm{a}}$ reduces to $\ket{\Psi_\mathrm{i}}$ in the limit of no errors yielding \mbox{$\mathcal{F}=1$}. Calculating this overlap but letting $\ket{\Psi_\mathrm{i}}$ have general temporal mode mismatch until the end of the calculation we obtain,
\begin{widetext}
\begin{eqnarray} \label{eq:Fidai}
\mathcal{F}&=& \bigg|\underbrace{\bra{0}^{\otimes m}\sum_{S'}\sum_{T'(S')}\bigg[\gamma_{S',T'}\prod_{i'=1}^{m} \hat{\mathcal{C}}\big(t_{S'_{i'}},\Delta_{T'(S_{i'}')}\big)\bigg]}_{\bra{\Psi_\mathrm{i}}} \underbrace{\sum_{S}\sum_{T(S)}\bigg[\gamma_{S,T}\prod_{i=1}^{m} \hat{\mathcal{C}}^{\dag}\big(t_{S_i},\Delta_{T(S_i)}\big) \bigg]\ket{0}^{\otimes m}}_{\ket{\Psi_\mathrm{a}}} \bigg|^2  \nonumber \\
&=& \bigg| \sum_{S',S,}\,\,\sum_{T'(S'),T(S)}\bigg[\gamma_{S',T'}\gamma_{S,T} \bra{0}^{\otimes m} \prod_{i'=1}^{m} \hat{\mathcal{C}}\big(t_{S'_{i'}},\Delta_{T'(S_{i'}')}\big) \prod_{i=1}^{m} \hat{\mathcal{C}}^{\dag}\big(t_{S_i},\Delta_{T(S_i)}\big) \ket{0}^{\otimes m} \bigg] \bigg|^2.
\end{eqnarray}
\end{widetext}
To simplify this expression further we use the formalism of second quantisation \cite{bib:berazin2012}, which describes how the indistinguishability of particles in quantum mechanics undergo symmetrisation. Here we use the exchange symmetry of the bosonic Fock states, which accounts for how each temporal photon annihilation operator $\hat{\mathcal{C}}\big(t_{S'_{i'}},\Delta_{T'(S_{i'}')}\big)$ overlaps with each temporal photon creation operator $\hat{\mathcal{C}}^{\dag}\big(t_{S_{i}}, \Delta_{T(S_i)}\big)$. Using bosonic exchange symmetry we sum over all $m!$ permutations of $\bigotimes_{i'=1}^{m} \hat{\mathcal{C}}\big(t_{S'_{i'}},\Delta_{S_{i'}'}\big) \bigotimes_{i=1}^{m} \hat{\mathcal{C}}^{\dag}\big(t_{S_i},\Delta_{S_i}\big)$. Then Eq. \ref{eq:Fidai} becomes,
\begin{widetext}
\begin{eqnarray} \label{eq:Fidai2}
\mathcal{F}&=& \Bigg| \sum_{S',S}\,\,\sum_{T'(S'),T(S)}\Bigg[\gamma_{S',T'}\gamma_{S,T}\sum_{\sigma} \bigg[\prod_{i=1}^{m} \bra{0}\hat{\mathcal{C}}\big(t_{S'_{\sigma_{i'}}},\Delta_{T'(S_{\sigma_{i'}}')}\big) \hat{\mathcal{C}}^{\dag}\big(t_{S_i},\Delta_{T(S_i)}  \big) \ket{0} \bigg] \Bigg] \Bigg|^2,
\end{eqnarray}
\end{widetext}
where $\sigma$ are the permutations over $m$ elements.

Finally, to calculate $\mathcal{F}$ we must find the wave packet simplification for \mbox{$\bra{0}\hat{\mathcal{C}}\big(t_{S'_{\sigma_{i'}}},\Delta_{T'(S_{\sigma_{i'}}')}\big) \hat{\mathcal{C}}^{\dag}\big(t_{S_i},\Delta_{T(S_i)} \ket{0}$}, which we perform in App. \ref{app:WavePacketSimp}. Using this result we obtain,
\begin{widetext}
\begin{eqnarray} \label{eq:Fidai3}
\mathcal{F} = \Bigg| \sum_{S',S}\,\,\sum_{T'(S'),T(S)}\Bigg[\gamma_{S',T'}\gamma_{S,T}\sum_{\sigma}\bigg[\prod_{i=1}^{m} \mathrm{exp}\bigg(-\frac{\big(\Delta_{T'(S'_{\sigma_{i'}})} -\Delta_{T(S_{i})}\big)^2}{4 c^2}\bigg) \bigg] \Bigg] \Bigg|^2.
\end{eqnarray}
\end{widetext}
Letting the ideal state $\ket{\Psi_\mathrm{i}}$ have no temporal shifts, \mbox{$\Delta \to 0$}, this reduces to,
\begin{widetext}
\begin{eqnarray} \label{eq:Fidai4}
\mathcal{F} &=& \Bigg| \sum_{S',S} \,\, \sum_{T'(S'),T(S)}\gamma_{S',T'}\cdot \gamma_{S,T} \cdot m! \cdot \mathrm{exp}\bigg(-\frac{m\Delta_{T(S_{i})}^2}{4 c^2}\bigg) \Bigg|^2.
\end{eqnarray}
\end{widetext}
This derivation assumes the width of all wave-packets remain the same, i.e the photons are identical up to a temporal displacement. The width of the wave-packets may broaden due to dispersion but under the relatively short lengths of fiber-loop required for small $m$ the effect of dispersion may be neglected; however this formalism may be easily modified to include dispersion by creating an operator that broadens the wave-packet width $c$ as a function of the length of the fiber the wave-packet has traversed.

Next we consider two types of mode-mismatch: non-ideal lengths of the inner loop, and time-jitter at the input source.

\subsection{Imperfect Inner Loop Length}

\begin{figure}[!htb]
\includegraphics[width=.8\columnwidth]{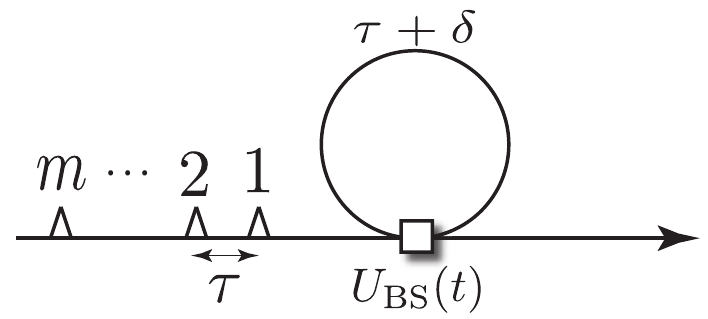}
\caption{The inner fiber-loop with an error $\delta$ in its intended length. Every time a pulse traverses the inner loop it is shifted from its expected temporal position by $\delta$, thereby reducing the Hong-Ou-Mandel visibility at $U_\mathrm{BS}$.} \label{fig:singleloopDeltaerror}
\end{figure}

Here we analyse errors in the MGDR fiber-loop architecture caused by a non-ideal length of inner fiber-loop as shown in Fig. \ref{fig:singleloopDeltaerror}. We let the length of the inner loop have some length \mbox{$\tau + \delta$}, where $\delta$ is the error in the intended length $\tau$ and may be positive or negative. Thus every photon that traverses the inner loop acquires a temporal shift of $\delta$ from its expected centre. We ignore imperfect lengths of the outer loop because every mode will traverse the outer loop an equal number of times creating a global temporal shift with no impact on interference at the central beamsplitter.

\begin{figure}[!htb]
\centering
\subfloat[Part 1][]
{\includegraphics[width=.8\columnwidth]{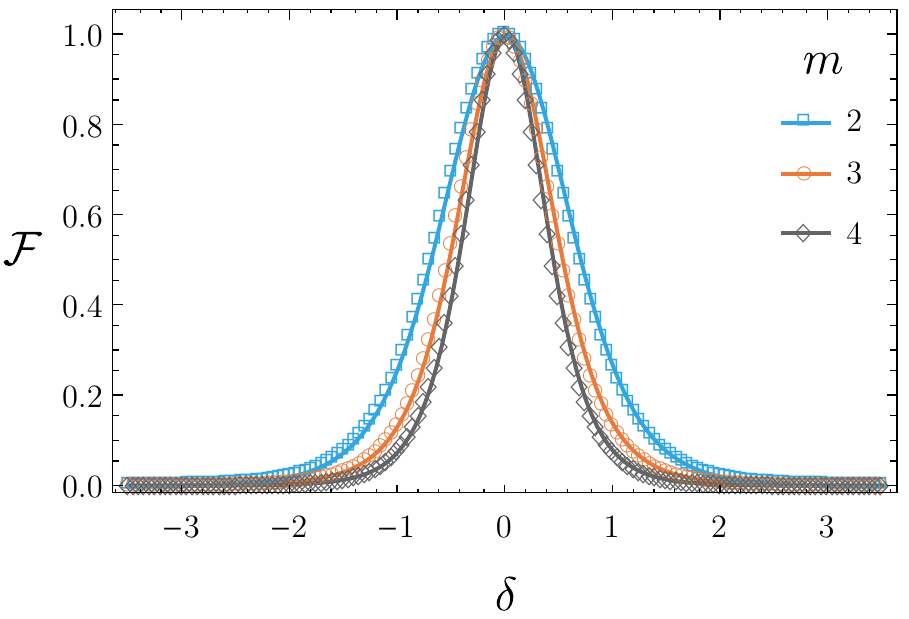} \label{fig:FidInnerLoopDelay}}\\
\subfloat[Part 2][]
{\includegraphics[width=.8\columnwidth]{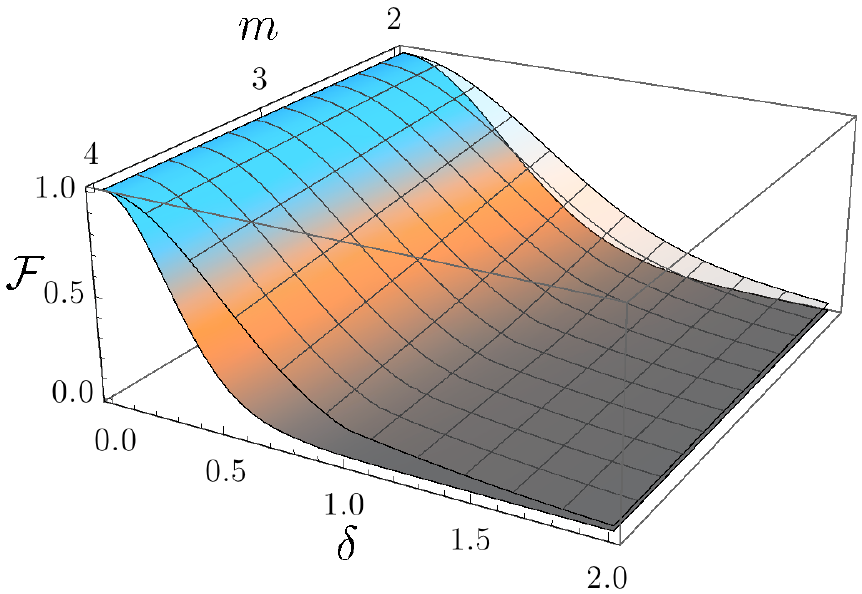} \label{fig:BestWorstFidJitter}}
\caption{(a) The average fidelity $\mathcal{F}$ between the ideal state $\ket{\Psi_\mathrm{i}}$ and the actual experimental state $\ket{\Psi_\mathrm{a}}$ versus the error in the intended length of the inner loop $\delta$. (b) The worst (bottom) and best (top) case fidelity $\mathcal{F}$ between the ideal state $\ket{\Psi_\mathrm{i}}$ and the actual experimental state $\ket{\Psi_\mathrm{a}}$ versus the error in the intended length of the inner loop $\delta$ and number of modes $m$. In (a) and (b) there are $m$ modes with one photon per mode and the data was obtained over 250 implementations each with a unique randomly generated unitary.}
\label{fig:Fidcombined1}
\end{figure}

The input state is given by Eq. \ref{eq:inputState} where \mbox{$\Delta_i=0\ \forall\ i$}. This models an ideal input state with no time-jitter or other errors in the source. To account for the unwanted time-delay $\delta$ we introduce the time-delay operator \mbox{$\hat{\mathrm{T}}(\delta)$},
\begin{equation}
\hat{\mathrm{T}}(\delta)\hat{\mathcal{B}}^{\dag}(t,\Delta)\hat{\mathrm{T}}^{\dag}(\delta)= \hat{\mathcal{B}}^{\dag}(t,\Delta+\delta),
\end{equation}
which acts only in region $\mathcal{B}$ -- the region inside the inner loop. This adds a small temporal displacement, not enough to confuse time-bins. Thus it affects $\Delta$ but not $t$. It has no effect on the mode-operators $\hat{\mathcal{A}}$ and $\hat{\mathcal{C}}$. Using the boundary conditions shown in the MGDR protocol, the first photon is coupled completely into the loop so it picks up a time-delay of $\delta$. Afterwards the pulse-train interacts at the beamsplitter described in Eqs. \ref{eq:BSA} and \ref{eq:BSB}, where it is sent into a superposition of regions $\mathcal{B}$ and $\mathcal{C}$. As the state evolves all amplitudes entering the inner loop (region $\mathcal{B}$) will acquire a time-shift of $\delta$. After the last mode traverses the inner loop the state is coupled completely out as per the MGDR protocol. 
The output state is given by,
\begin{widetext}
\begin{eqnarray} \label{}
\ket{\Psi}_{\mathrm{out}} &=& \left[\prod_{i'=1}^{t}\hat{\mathrm{T}}(\delta)\hat{U}_{\mathrm{BS}}(i')\right] \cdot \left[\prod_{i=1}^{m} \frac{1}{\sqrt{k_i!}}\hat{\mathcal{A}}^{\dag}(t_{i},\Delta_i)^{k_i}\right]\cdot \left[\prod_{i'=1}^{t}\hat{\mathrm{T}}(\delta)\hat{U}_{\mathrm{BS}}(i')\right]^\dag \ket{0}^{\otimes m},
\end{eqnarray}
\end{widetext}
where we have inserted the time-delay operator appropriately in Eq. \ref{eq:outevo}. Fig. \ref{fig:FidInnerLoopDelay} shows how the fidelity $\mathcal{F}$ scales with $m$, $n$, and $\delta$ and Fig. \ref{fig:BestWorstFidJitter} shows the worst- and best-case fidelities, where we have searched over switching sequences.

\subsection{Time-jitter from Input Source}
\begin{figure}[!htb]
\centering
\subfloat[Part 1][]
{\includegraphics[width=.8\columnwidth]{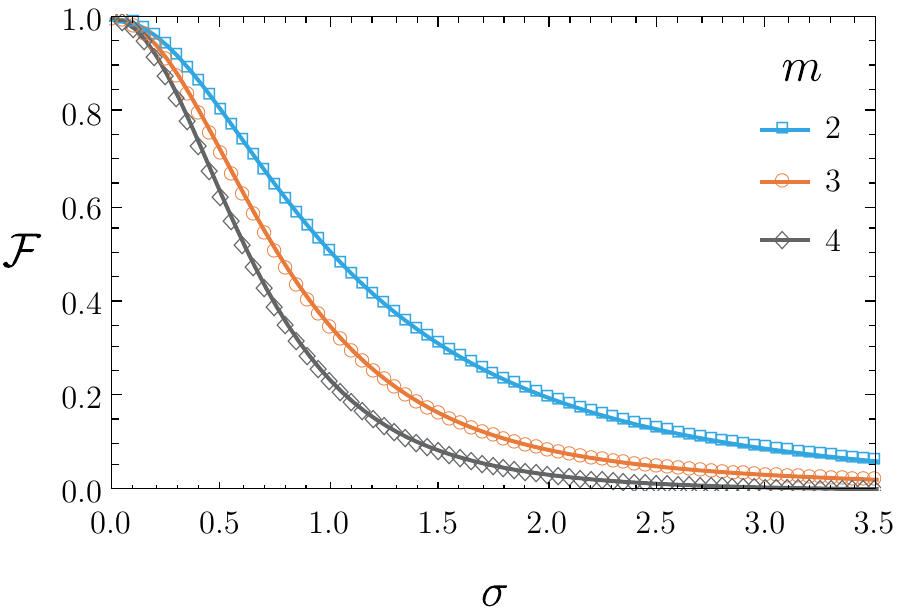} \label{fig:FidJitter}}\\
\subfloat[Part 2][]
{\includegraphics[width=.8\columnwidth]{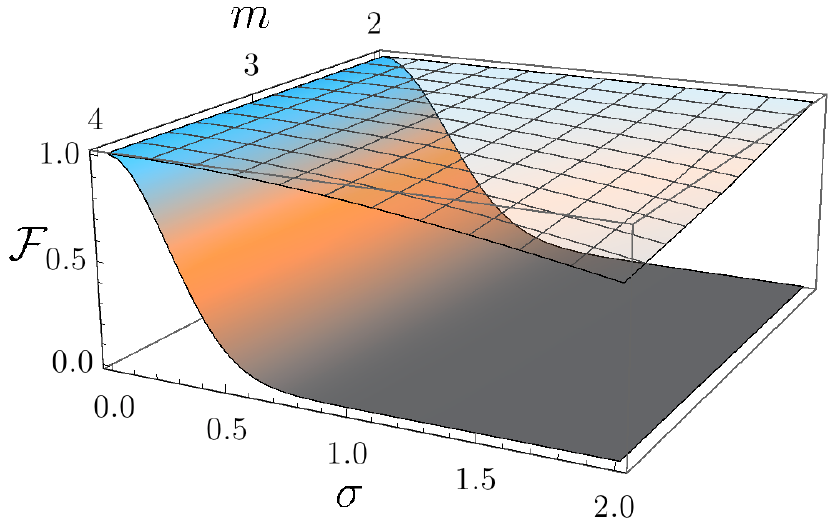} \label{fig:BestWorstFidJitter}}
\caption{(a) The fidelity $\mathcal{F}$ between the ideal state $\ket{\Psi_i}$ and the actual experimental state $\ket{\Psi_a}$ with random time-jitter in the input source versus modes $m$ and standard deviation $\sigma$ with no fiber length error $\delta=0$. (b) The worst (bottom) and best (top) case fidelity $\mathcal{F}$ between the ideal state $\ket{\Psi_\mathrm{i}}$ and the actual experimental state $\ket{\Psi_\mathrm{a}}$ with time-jitter. In (a) and (b) there is one photon per mode, the data was averaged over 250 implementations each with a unique randomly generated unitary, and the time-jitter was drawn from the normal distribution.}
\label{fig:Fidcombined2}
\end{figure}

A major source of error in the time-bin architecture is time-jitter of the input source. Ideally each mode will be separated by time $\tau$ but in reality non-ideal sources will randomly shift modes from their desired centre of time $t_i$ in mode $i$. To model time-jitter we let the temporal shift of input mode $i$ be a Gaussian random variable $\epsilon_i$ drawn from the normal distribution,
\begin{eqnarray}
\mathcal{N}_i(\epsilon_i)=\frac{1}{\sigma\sqrt{2\pi}}\mathrm{exp}\left(-\frac{(\epsilon_i-t_i)^2}{2 \sigma^2}\right),
\end{eqnarray}
centered in mode $i$ at time $t_i$ and with a standard deviation of $\sigma$. The input state of Eq. \ref{eq:inputState} becomes,
\begin{equation}
\ket{\psi_\mathrm{in}}= \bigotimes_{i=1}^{m}\frac{1}{\sqrt{k_i!}} \hat{\mathcal{A}}^{\dag}\big(t_i, \epsilon_i\big)^{{k_i}}\ket{0}_i.
\end{equation}
We assume that the shifts caused by time-jitter are much less than the time-bin separation $\tau$, such that the probability of time-bin confusion remains negligible, i.e. \mbox{$\mathcal{N}(\epsilon_i) \ll \tau$}. Fig. \ref{fig:FidJitter} shows how the fidelity $\mathcal{F}$ scales with $m$, $n$, and $\sigma$. Fig. \ref{fig:BestWorstFidJitter} shows the worst- and best-case $\mathcal{F}$, searching over many switching sequences.

%Section
\section{Conclusion}
In this work we have analysed sources of error in the Motes, Gilchrist, Dowling \& Rohde fiber-loop architecture for implementing \BS. Specifically we have analysed loss and mode-mismatch. In the loss analysis we examined how lossy fibers and switches affect the operation of the architecture in both the inner and outer loops. We found that loss causes an asymmetric bias in the desired unitary, unique to a temporally implemented unitary transformation. That is, even upon post-selection the operation of the device is erroneous. Additionally, like all linear optical architectures, our scheme has exponential dependence on loss, thereby reducing the post-selection success probability of detecting all $n$ photons. In the mode-mismatch analysis we analysed only the inner loop since no interference occurs in the outer loop. We examined two types of mode-mismatch including an imperfect length of fiber in the inner loop, and time-jitter of the photon source. This analysis provides a guideline for future experimental implementations, to provide insight into how such a device might realistically behave in the presence of loss and mode-mismatch, the two dominant error mechanisms affecting this protocol.

% ACKNOWLEDGEMENTS
\begin{acknowledgements}
KRM and AG acknowledges the Australian Research Council Centre of Excellence for Engineered Quantum Systems (Project number CE110001013). PPR acknowledges support from Lockheed Martin. JPD acknowledges support from the Air Force Office of Scientific Research, the Army Research Office, The National Science Foundation, and Northrop-Grumman. We acknowledge Professor Geoff Pryde for helpful discussions. 
\end{acknowledgements}
\vspace{\stretch{-11}}

% BIBLIOGRAPHY
\bibliography{bibliography}

%Appendix
\appendix

\section{Intuitive Example of \\Loop Bias due to Loss} \label{app:MultipleLoopBiasExample}
An example of how $\hat{U}$ becomes biased is explained here. Let's consider two examples of a two-mode pulse-train --- a single inner loop and two inner loops. 

\subsection{One Loop}
The first mode can can exit the first output mode by traversing the inner loop once. Here it picks up loss due to the middle switch twice ${\eta_s}^2$, and loss due to the inner loop fiber once $\eta_f$, obtaining a net loss of \mbox{${\eta_s}^2\eta_f$}. The first mode can exit the second output mode by traversing the inner loop twice. In this case it obtains a net loss of \mbox{${\eta_s}^3{\eta_f}^2$}. A similar analysis can be performed for the other combinations. Then, we can write the loss amplitudes corresponding to the input (rows) and output (columns) modes in matrix form as, 
\begin{equation} 
\mathcal{\hat{L}} = \left(\begin{array}{cc}
{\eta_s}^2\eta_f & {\eta_s}^3{\eta_f}^2  \\
\eta_s &{\eta_s}^2\eta_f 
\end{array}\right) = \eta_s \left(\begin{array}{cc}
\eta & {\eta}^2  \\
1 &\eta 
\end{array}\right),
\end{equation}
where $\eta=\eta_s\eta_f$ and observe the bias accumulating in this input to output map. The net input-to-output mapping of amplitudes is given by taking the element-wise product of this loss matrix with the ideal unitary, \mbox{$\mathcal{\hat{L}}\circ \hat{U}$}, thereby leaving us with a biased map.

\subsection{Two Loops}
A similar analysis as above but following the paths for two consecutive applications of the inner loop (i.e one traversal of the outer loop), we find the input-to-output loss matrix to be,
\begin{equation} 
\mathcal{\hat{L}} = \left(\begin{array}{cc}
{\eta_s}^4{\eta_f}^2 & {\eta_s}^5{\eta_f}^3  \\
{\eta_s}^3\eta_f &{\eta_s}^4{\eta_f}^2 
\end{array}\right) = {\eta_s}^2 \eta \left(\begin{array}{cc}
\eta & {\eta}^2  \\
1 &\eta 
\end{array}\right). 
\end{equation}
where we have ignored the losses due to the outer loop as it yields an overall normalisation factor that does not bias $\hat{U}$. As we can see, for each iteration of the inner loop $\hat{U}$ accumulates more loss, with a decreasing overall success probability, but the amount of skew in the matrix remains the same.

\section{Wave-packet Simplifications}\label{app:WavePacketSimp}

In this section we derive the overlap of two photons with temporal creation operator \mbox{$\hat{\mathcal{A}}^{\dag}(t,\Delta)$} as per Eq. \ref{eq:TPCO} and temporal density function $\psi(t,\Delta)$. For the purpose of this work we assume that the temporal spacing $\tau$ between each mode is much larger than the width of the wave packet $c$ such that the overlap of our temporal photons in different time-bins is negligible,
\begin{equation}
\bra{0}\hat{\mathcal{A}}(t,\Delta)\hat{\mathcal{A}}^{\dag}(t',\Delta')\ket{0}=0,
\end{equation}
for $t \neq t'$. 

For photons in the same time-bin and allowing for arbitrary temporal-shifts, the overlap is,
\begin{eqnarray}
&\ &\bra{0}\hat{\mathcal{A}}(t,\Delta')\hat{\mathcal{A}}^{\dag}(t,\Delta)\ket{0} \nonumber \\
&=&\bigg(\bra{0}\int_{-\infty}^{\infty}\psi^{*}(x'-t-\Delta')\hat{a}(x')dx'\bigg)\nonumber\\
&\times& \bigg(\int_{-\infty}^{\infty}\psi(x-t-\Delta)\hat{a}^{\dag}(x)dx \ket{0} \bigg) \nonumber \\
&=& \int_{-\infty}^{\infty}\int_{-\infty}^{\infty}\psi^{*}(x'-t-\Delta')\psi(x-t-\Delta) \nonumber \\
&\times& \underbrace{\bra{0}\hat{a}(x')\hat{a}^{\dag}(x)\ket{0}}_{\delta_{x',x}}dx'dx \nonumber \\
&=& \int_{-\infty}^{\infty} \psi^{*}(x-\Delta')\psi(x-\Delta)dx \nonumber \\
&=& e^{-\frac{(\Delta' -\Delta )^2}{4 c^2}}.
\end{eqnarray}

For ideal states where \mbox{$\Delta'=\Delta=0$}, we notice that \mbox{$\mathcal{F}=1$}, as expected when there is no mode-mismatch. We use these results for simplifying Eq. \ref{eq:Fidai3} in our analysis of mode-mismatch.

\end{document}